\documentclass[12pt]{article}
\usepackage{epsfig,amsmath,amssymb}
\usepackage{graphicx,psfrag} %,float,color  a4paper,
\newcommand{\tinynot}{\mbox{\tiny $0$}}

\newcommand{\geff}{g_{\mbox{\tiny eff}}}
\newcommand{\be}{\begin{equation}}
\newcommand{\ee}{\end{equation}}
\newcommand{\bea}{\begin{eqnarray}}
\newcommand{\eea}{\end{eqnarray}}
\newcommand{\bA}{\begin{array}}
\newcommand{\eA}{\end{array}}
\newcommand{\bc}{\begin{center}}
\newcommand{\ec}{\end{center}}

\newcommand{\ra}{\rightarrow}
\newcommand{\del}{\partial}

\newcommand{\ie}{{\it i.e.}}
\newcommand{\eg}{{\it e.g.}}

\begin{document}

%\ifeprint
%\fi

\begin{titlepage}
%\vspace{30mm}

\bc

%\hfill  {TIFR/TH/09-12} \\
\hfill % {\tt arXiv:0909.4731 [hep-th]} 
\\         [22mm]
%%X\vfill

{\Huge Hyperscaling violation\\
[2mm] 
and the shear diffusion constant}
\vspace{16mm}

Kedar S. Kolekar, Debangshu Mukherjee, K. Narayan\\
\vspace{3mm}
{\small \it Chennai Mathematical Institute, \\ }
{\small \it SIPCOT IT Park, Siruseri 603103, India.\\ }
%{\small Email: \ narayan@cmi.ac.in}\\

\ec
\medskip
\vspace{40mm}

\begin{abstract}
We consider holographic theories in bulk $(d+1)$-dimensions with
Lifshitz and hyperscaling violating exponents $z,\theta$ at finite
temperature. By studying shear gravitational modes in the near-horizon
region given certain self-consistent approximations, we obtain the
corresponding shear diffusion constant on an appropriately defined
stretched horizon, adapting the analysis of Kovtun, Son and Starinets. 
For generic exponents with $d-z-\theta>-1$, we find that the diffusion
constant has power law scaling with the temperature, motivating us to
guess a universal relation for the viscosity bound. When the exponents
satisfy $d-z-\theta=-1$, we find logarithmic behaviour. This relation
is equivalent to $z=2+d_{eff}$ where $d_{eff}=d_i-\theta$ is the
effective boundary spatial dimension (and $d_i=d-1$ the actual
spatial dimension).  It is satisfied by the exponents in
hyperscaling violating theories arising from null reductions of highly
boosted black branes, and we comment on the corresponding analysis in
that context.
\end{abstract}

\end{titlepage}

%\newpage 
%{\tiny %footnotesize
%\begin{tableofcontents}
%\end{tableofcontents}
%}

%\vspace{2mm}

\section{Introduction and summary} \label{adspwreview}

The viscosity bound \cite{Kovtun:2004de} is a universal feature of
large families of strongly coupled quantum field theories arising in
investigations using holography \cite{AdSCFT}.  The shear viscosity
$\eta$ satisfies ${\eta\over s} = {1\over 4\pi}$ for a wide variety of
theories, $s$ being the entropy density.  A slightly different
approach to studying hydrodynamics and viscosity was studied in
\cite{Kovtun:2003wp}. It was observed that metric perturbations
governing diffusive charge and shear modes in the near horizon region
of the relevant dual black branes simplify allowing a systematic
expansion there, resulting in a diffusion equation on a stretched
horizon with universal behaviour for the diffusion constant. This is
akin to the membrane paradigm \cite{Thorne:1986iy} for black branes,
the horizon exhibiting diffusive properties. This approach does not
directly assume any holographic duality per se, although it is
consistent with holographic results \eg\
\cite{Policastro:2001yc,Son:2002sd}\ (see \eg\ \cite{Son:2009zzd} for
a review of these aspects of hydrodynamics).

In recent years, nonrelativistic generalizations of gauge/gravity
duality have been studied extensively. An interesting class of
non-relativistic theories exhibits so-called hyperscaling violation.
The gravity duals are conformal to Lifshitz spacetimes 
\cite{Kachru:2008yh,Taylor:2008tg}. These hyperscaling violating 
spacetimes arise in effective Einstein-Maxwell-Dilaton theories \eg\ 
\cite{Gubser:2009qt,Cadoni:2009xm,Goldstein:2009cv,Charmousis:2010zz,
Perlmutter:2010qu,Gouteraux:2011ce,Iizuka:2011hg,Ogawa:2011bz,
Huijse:2011ef,Dong:2012se,Alishahiha:2012cm,Bhattacharya:2012zu}. 
Certain gauge/string realizations of these arise in null $x^+$-reductions 
of $AdS$ plane waves \cite{Narayan:2012hk,Narayan:2012wn}, which are 
large boost, low temperature limits \cite{Singh:2012un} of boosted 
black branes \cite{Maldacena:2008wh}. Various aspects of Lifshitz 
and hyperscaling violating holography appear in \eg\ 
\cite{Dong:2012se,Ross:2011gu,Christensen:2013rfa,Chemissany:2014xsa,
Hartong:2015wxa,Taylor:2015glc}. Some of these exhibit novel scaling 
for entanglement entropy \eg\ \cite{Ogawa:2011bz,Huijse:2011ef,
Dong:2012se}: the string realizations above reflect this 
\cite{Narayan:2012ks,Narayan:2013qga,Mukherjee:2014gia,Narayan:2015lka}, 
suggesting corresponding regimes in the gauge theory duals exhibiting 
this scaling.

It is of interest to study hydrodynamic behaviour in these
nonrelativistic generalizations of holography: previous investigations
include \eg\ \cite{Pang:2009wa,Hoyos:2013qna,Sadeghi:2014zia,
Roychowdhury:2014lta,Kiritsis:2015doa,Kuang:2015mlf,Blake:2016wvh}.
In this paper, we study the shear diffusion constant in bulk
$(d+1)$-dimensional hyperscaling violating theories (\ref{hvmetric})
with $z,\theta$ exponents adapting the membrane-paradigm-like analysis
of \cite{Kovtun:2003wp}. As in \cite{Kovtun:2003wp}, we map the
diffusion of shear gravitational modes on a stretched horizon to
charge diffusion in an auxiliary theory obtained by compactifying one
of the $d_i$ boundary spatial dimensions exhibiting translation 
invariance. This gives a near horizon expansion for perturbations with 
modifications involving $z,\theta$. We find (sec.~2) that for generic 
exponents with $d-z-\theta>-1$, the shear diffusion constant is\ 
${\cal D} = {r_0^{z-2}\over d-z-\theta-1}$~, \ie\ power-law scaling
(\ref{DT(2-z)/z}) with the temperature $T\sim r_0^z$. Studying various
special cases motivates us to guess (\ref{eta/s,DT(2-z)/zHV}), \ie\
$\# {\cal D} T^{{2-z\over z}} = {1\over 4\pi}$\ where $\#$ is some
$(d,z,\theta)$-dependent constant, suggesting that ${\eta\over s}$ has
universal behaviour. The condition $z<2+d_i-\theta$ representing this
universal sector appears related to requiring standard quantization
from the point of view of holography. It would be interesting to
understand the hydrodynamics and viscosity here more systematically.

When the exponents satisfy\ $d-z-\theta= -1$, the diffusion constant
exhibits logarithmic behaviour (sec.~2.2), suggesting a breakdown of
some sort in this analysis. This condition appears compatible however
with various known constraints on the exponents: it would be
interesting to understand this from other considerations.  The
exponents arising in null reductions of AdS plane waves or highly 
boosted black branes \cite{Narayan:2012hk,Singh:2012un,Narayan:2012wn} 
satisfy this condition, which can be written as $z=2+d_{eff}$. It is
interesting to note that highly boosted black branes (or $AdS$ plane
waves) give rise upon $x^+$-reduction to hyperscaling violating
theories with $z, \theta$ exponents leading to novel entanglement
scaling, as well as the condition $z=2+d_{eff}$ here. The two appear
independent however: the entanglement entropy stems from $d_i-1\leq
\theta\leq d_i$ and does not depend on $z$ while the relation here
involves both $z, \theta$.  We discuss (sec.~3) the corresponding
picture of hydrodynamics that is likely to arise in the null reduction
by mapping the perturbations accordingly. Details of the diffusion
analysis appear in the Appendix.

\section{Shear diffusion on the stretched horizon}

We are considering nonrelativistic holographic backgrounds described by 
a $(d+1)$-dim hyperscaling violating metric at finite temperature,
\be\label{hvmetric}
ds^2 =  r^{2\theta/d_i}\left(-\frac{f(r)}{r^{2z}}dt^2 + 
\frac{dr^2}{r^2f(r)}+\frac{\sum_{i=1}^{d_i} dx_{i}^{2}}{r^2}\right) ,
\qquad  d_i=d-1 ,\qquad d_{eff}=d_i-\theta\ ,
\ee
where $f(r) = 1-(r_0r)^{d+z-\theta-1}$.\ $d_i$ is the boundary spatial 
dimension while $d_{eff}$ is the effective spatial dimension governing 
various properties of these theories, for instance the entropy density 
$s\sim T^{d_{eff}/z}$. The temperature of the boundary field theory 
(\ie\ Hawking temperature of the black hole) is
\begin{equation}
\label{hvtemperature}
T = \frac{(d+z-\theta-1)}{4\pi}~r_0^z\ .
\end{equation}
These spacetimes are conformal to Lifshitz spacetimes 
\cite{Kachru:2008yh,Taylor:2008tg}, and exhibit\
$t \rightarrow \lambda^z t, \ x_i \rightarrow \lambda x_i$,\ 
$r \rightarrow \lambda r, \ ds \rightarrow \lambda^{\theta/(d-1)}ds$.
They arise in Einstein-Maxwell-Dilaton theories and are sourced by 
gauge fields and scalars. The window $d_i-1\leq \theta\leq d_i$ 
shows novel scaling for entanglement entropy 
\cite{Ogawa:2011bz,Huijse:2011ef,Dong:2012se}: these arise in the 
string realizations \cite{Narayan:2012hk,Singh:2012un,Narayan:2012wn}, 
with entanglement entropy studies in \cite{Narayan:2012ks,Narayan:2013qga,
Mukherjee:2014gia,Narayan:2015lka}. The null energy conditions 
following from (\ref{hvmetric}) constrain the exponents, giving
\be\label{energyconds}
(d-1-\theta) \big((d-1)(z-1)-\theta\big) \geq 0\ ,\qquad\quad 
(z-1)(d-1+z-\theta) \geq 0\ .
\ee

We want to study the diffusion of shear gravitational modes in these
backgrounds as a way of studying shear viscosity. In
\cite{Kovtun:2003wp}, Kovtun, Son and Starinets formulated charge and
shear diffusion for black brane backgrounds in terms of
long-wavelength limits of perturbations on an appropriately defined
\emph{stretched horizon}, the broad perspective akin to the membrane
paradigm \cite{Thorne:1986iy}. Their analysis, which is quite general,
begins with a background metric of the form
\begin{equation}\label{genmetric}
ds^2 = G_{\mu\nu} dx^\mu dx^\nu = 
G_{tt}(r)dt^2+G_{rr}(r)dr^2+G_{xx}(r)\sum_{i=1}^{d-1}dx_i^2\ ,
\end{equation}
which includes the hyperscaling violating backgrounds \eqref{hvmetric} 
as a subfamily. Charge difffusion of a gauge field perturbation $A_\mu$ 
in the background \eqref{genmetric} is encoded by the charge diffusion 
constant $D$, defined through Fick's Law $j^{i}=-D\partial_{i}j^{t}$, 
where the 4-current $j^{\mu}$ is defined on the stretched horizon 
$r=r_h$ (with $n$ the normal) as $j^{\mu}= n_{\nu}F^{\mu \nu}|_{r=r_h}$.
Then current conservation $\del_\mu j^\mu = 0$ leads to the diffusion 
equation $\del_tj^t = -\del_ij^i = D \del_i^2 j^t$, with $D$ the 
corresponding diffusion constant.  Fick's law in turn can be shown to
apply if the stretched horizon is localized appropriately with regard
to the parameters $\Gamma, q, T$.  Translation invariance along $x\in
\{x_i\}$ allows considering plane wave modes for the perturbations\
$\propto e^{-\Gamma t + iqx}$, where $\Gamma$ is the typical time
scale of variation and $q$ the $x$-momentum. In the IR regime, the
modes vary slowly: this hydrodynamic regime is a low frequency, 
long wavelength regime.

The diffusion of shear gravitational modes can be mapped to charge
diffusion \cite{Kovtun:2003wp}: under Kaluza-Klein compactification of
one of the directions along which there is translation invariance,
tensor perturbations in the original background map to vector
perturbations on the compactified background.
Here we carry out a similar analysis for the shear diffusion constant 
in the backgrounds (\ref{hvmetric}), adapting \cite{Kovtun:2003wp} to 
the present context. We 
turn on the metric fluctuations $h_{xy}$ and $h_{ty}$ ($x\equiv x_1$,\
$y\equiv x_2$) around \eqref{genmetric}, depending only on $t, r, x$, 
\ie\ $h_{ty}=h_{ty}(t,x,r) , \ h_{xy}=h_{xy}(t,r,x)$.\
Other fluctuation modes can be consistently set to zero.
There is translation invariance along the $y$-direction: thus after a
$y$-compactification, the modes $h_{xy}$ and $h_{ty}$ become
components of a $U(1)$ gauge field in the dimensionally reduced 
$d$-dim spacetime. The components are given by 
\be\label{yCompax}
{g}_{\mu \nu} = {G}_{\mu \nu}(G_{xx})^{\frac{1}{d-2}} \qquad 
[\mu, \nu = 0,\ldots, d-1];\qquad
A_0 = (G_{xx})^{-1}h_{ty}\ ,\quad A_x = (G_{xx})^{-1}h_{xy}\ ,
\ee
where $G_{\mu \nu}$ is the metric given by \eqref{genmetric}. A part of 
the gravitational action contains the Maxwell action with an 
$r$-dependent coupling constant,\
$\sqrt{-G}{R}\ \ra\ -\frac{1}{4}\sqrt{-g}F_{\alpha \beta}F_{\gamma \delta}
{g}^{\alpha \gamma}{g}^{\beta \delta}(G_{xx})^{\frac{d-1}{d-2}}$.\
%\ee%\begin{equation*} %\end{equation*}
The gauge field equations following from the action are
\be\label{gaugefieldaction}
\partial_{\mu}\Big(\frac{1}{\geff^2}\sqrt{-g}F^{\mu \nu}\Big)=0\ ,
\qquad\quad {1\over \geff^2} = G_{xx}^{\frac{d-1}{d-2}}\ ,
\ee
where we have read off the $r$-dependent $\geff$ from the compactified 
action. Analysing these Maxwell equations and the Bianchi identity 
assuming gauge field ansatze\ $A_{\mu}=a_{\mu}(r)e^{-\Gamma t+iqx}$ and 
radial gauge $A_r=0$ as in \cite{Kovtun:2003wp} shows interesting 
simplifications in the near-horizon region.
When $q=0$, these lead to\ 
$\partial_r \big(\frac{\sqrt{-g}}{\geff^2}g^{rr}g^{tt}\partial_rA_t\big)=0$.
We impose the boundary condition that the gauge fields vanish at 
$r=r_c\sim 0$.
As in \cite{Kovtun:2003wp}, for $q$ nonzero but small, we assume an 
ansatz for $A_t$ as a series expansion in $\frac{q^2}{T^{2/z}}$
\begin{equation}\label{ansatz1}
\begin{aligned}
&\qquad A_t = A_t^{(0)}+ A_t^{(1)}+ \ldots\ , \qquad\qquad\quad 
A_t^{(1)} = O\Big({q^2\over T^{2/z}}\Big)\ ,\\
&A_t^{(0)} = Ce^{-\Gamma t+iqx}\int_{r_c}^{r}dr' 
\frac{g_{tt}(r')g_{rr}(r')}{\sqrt{-g(r')}}\cdot \geff^2(r')\ 
= Ce^{-\Gamma t+iqx}\int_{r_c}^{r}dr'\ 
\frac{G_{tt}(r')G_{rr}(r')}{G_{xx}(r')\sqrt{-G(r')}}\ ,
\end{aligned}
\end{equation}
using (\ref{yCompax}), (\ref{gaugefieldaction}), with $C$ some constant. 
Making a second assumption
\begin{equation}\label{ansatz2}
|\partial_t A_x| \ll |\partial_x A_t|
\end{equation}
as in \cite{Kovtun:2003wp}, the gauge field component $A_x$, using the 
$A_t$ solution, becomes
\begin{equation}\label{axsoln}
\begin{aligned}
&\qquad\qquad\qquad A_x = A_x^{(0)}+A_x^{(1)}+ \ldots\ , \\
& A_x^{(0)} = -\frac{i\Gamma}{q}Ce^{-\Gamma t+iqx}\int_{r_c}^{r}dr' 
\frac{g_{xx}(r')g_{rr}(r')}{\sqrt{-g(r')}}\cdot \geff^{2}(r') 
= -\frac{i\Gamma}{q}Ce^{-\Gamma t+iqx}\int_{r_c}^{r}dr' 
\frac{G_{rr}(r')}{\sqrt{-G(r')}}\ ,
\end{aligned}
\end{equation}
again as a series expansion. As for $A_t$, we impose the 
boundary condition $A_x\ra 0$ as $r\ra r_c\sim 0$. 
In Appendix \ref{appendix1}, \ref{analysisspecial}, we show that the 
above series expansions are self-consistent provided certain conditions 
hold on the location $r_h$ of the stretched horizon and the parameters 
$q, \Gamma$ and $T$ (equivalently $r_0$).\ This enables us to define 
Fick's law on the stretched horizon, and thereby the diffusion equation. 
The shear diffusion constant then becomes
\begin{equation}
\label{chargediff}
{\cal D} =\frac{\sqrt{-g(r_h)}}{\geff^{2}(r_h)g_{xx}(r_h)
\sqrt{-g_{tt}(r_h)g_{rr}(r_h)}}\int_{r_c}^{r_h} dr 
\frac{-g_{tt}(r)g_{rr}(r)\geff^{2}(r)}{\sqrt{-g(r)}}\ ,
\end{equation}
where $r_c$ is the location of the boundary, and we are evaluating 
${\cal D}$ at the stretched horizon.
The leading solutions $A_{t,x}^{(0)}$ and ${\cal D}$ depend on the 
exponents: we analyse this separately below.

\subsection{Shear diffusion constant:\ \ $d-z-\theta > -1$\ \ (or\ $z<2+d_{eff}$)}

Using \eqref{hvmetric}, the expression (\ref{ansatz1}) for $A_t^{(0)}$ becomes
\be\label{At0soln}
A_t^{(0)} =\ Ce^{-\Gamma t+iqx}\int_{r_c}^r dr\ r^{d-z-\theta}\ .
\ee
For generic values 
\be\label{genExp}
d-z-\theta > -1\ ,
\ee
the leading solution (\ref{At0soln}) for $A_t^{(0)}$ has power law behaviour
\be\label{At0soln2}
A_t^{(0)} = \frac{C}{d-z-\theta+1} e^{-\Gamma t+iqx} 
\ r^{d-z-\theta+1}\ .
\ee
We expect that the hyperscaling violating phase breaks down close to 
the boundary at $r_c$: for our purposes, strictly speaking we will only 
require that the horizon is well-separated from the boundary, \ie\ 
$r_0r_c\ll 1$, or equivalently that the temperature is sufficiently 
below the ultraviolet cutoff in the theory.
Thus the condition $d-z-\theta>-1$ arises from the boundary condition 
$A_t^{(0)}=0$ at $r= r_c\sim 0$. This includes various subfamilies of 
hyperscaling violating metrics that arise in gauge/string realizations, 
\eg\ through dimensional reductions of nonconformal $Dp$-branes 
\cite{Dong:2012se} for $p\leq 4$\ (here $\theta=p-{9-p\over 5-p} ,\ 
d_i=p$).  %as well as some of the spacetimes in \cite{Narayan:2012hk}. 
The case of $d-z-\theta=-1$, arises in the reductions of various 
D-brane plane waves \cite{Narayan:2012hk} \cite{Singh:2012un} 
\cite{Narayan:2013qga}: here the leading solution has logarithmic 
behaviour, as we describe later.

From (\ref{hvmetric}), the condition (\ref{genExp}) can be written as\
$d_i-\theta - z + 2 > 0$,\ or\ $z < 2 + d_{eff}$, \ie\ the Lifshitz
exponent is not too high. For $z>2+d_{eff}$, it appears that the
perturbations (\ref{At0soln}) do not die far from the horizon. For
$z=1$, this gives $\theta>d_i+1$ which arises \eg\ from the reduction
of $D6$-branes: in such cases, it would appear that gravity does not
decouple and the asymptotics is not well-defined. It would be
interesting to understand the condition (\ref{genExp}) better from
other considerations \eg\ holography \cite{Ross:2011gu,
Christensen:2013rfa,Chemissany:2014xsa,Hartong:2015wxa,Taylor:2015glc}:\ 
more comments appear later.

The leading solution for $A_x$ likewise is
\be\label{Ax0soln}
A_x^{(0)} = -\frac{i\Gamma}{q}Ce^{-\Gamma t+iqx}
\frac{r_0^{\theta+1-d-z}}{\theta+1-d-z}
\log\left(1-(r_0r)^{d+z-\theta-1}\right)\ .
\ee
Self-consistency of \eqref{ansatz1}, (\ref{axsoln}), (\ref{ansatz2}),
when $d-z-\theta>-1$ leads to 
\begin{equation}\label{combinedcriteria}
e^{-\frac{T^{2/z}}{q^2}} \ll \frac{\frac{1}{r_0}-r_h}{\frac{1}{r_0}} \ll 
\frac{q^2}{T^{2/z}} \ll 1\ .
\end{equation}
This means that the stretched horizon has to be sufficiently close to 
the horizon (to $O(q^2)$) but not exponentially close to it. These 
conditions can be simultaneously satisfied in a self-consistent manner 
as we discuss in Appendix \ref{appendix1}, adapting \cite{Kovtun:2003wp}.

Using (\ref{yCompax}), (\ref{gaugefieldaction}), the shear diffusion 
constant (\ref{chargediff}) becomes
\begin{equation}%\begin{aligned}
\label{maindiffeqn}
\mathcal{D} = \frac{\sqrt{-G(r_h)}}{\sqrt{-G_{tt}(r_h)G_{rr}(r_h)}}
\int_{r_c}^{r_h}dr \frac{-G_{tt}(r)G_{rr}(r)}{G_{xx}(r)\sqrt{-G(r)}}\
=\ {1\over r_h^{d-\theta-1}}\int_{r_c}^{r_h} r^{d-z-\theta} dr \ .
%\end{aligned}
\end{equation}
Thus, for a hyperscaling violating theory with $d-z-\theta>-1$, we obtain  
\begin{equation}\label{Dhvgen}
\mathcal{D}\ = \ {r_h^{2-z}\over d-z-\theta+1}\ \simeq\ 
{r_0^{z-2}\over d-z-\theta+1}\ +\ O(q^2)\ ,
\end{equation}
where we have dropped the contribution in the integral from $r_c$ since 
the UV scale $r_c\ll r_h$ is well-separated from the horizon 
scale. The diffusion constant in (\ref{maindiffeqn}), (\ref{Dhvgen}), 
is evaluated at the stretched horizon $r_h$: however 
$r_h\sim {1\over r_0} + O(q^2)$ so that to leading order ${\cal D}$ is
evaluated at the horizon ${1\over r_0}$. 
It is interesting that $\theta$ cancels in the $r_0$-dependent terms 
in ${\cal D}$, which is essentially the ratio of $A_t$ to a field 
strength component (both of which have nontrivial $\theta$ dependence).

In the present hyperscaling violating case, we have seen that\ 
$T \sim r_0^{z}$\ and\ $\mathcal{D} \sim r_0^{z-2}$\ so the product 
$\mathcal{D}T \sim r_0^{2(z-1)}$\ is not dimensionless. 
Using (\ref{hvtemperature}), we have
\be\label{DT(2-z)/z}
{\cal D} = {1\over d-z-\theta+1} 
\Big({4\pi\over d+z-\theta-1}\Big)^{{z-2\over z}}\ 
T^{{z-2\over z}}\ ,
\ee
as the scaling with temperature $T$ of the leading diffusion constant 
(\ref{Dhvgen}). See also \eg\ \cite{Pang:2009wa,Cremonini:2011ej,
Sadeghi:2014zia,Roychowdhury:2014lta,Kuang:2015mlf,Blake:2016wvh} for 
previous investigations including via holography.

\subsubsection{Comments on ${\eta\over s}$} 

We now make a few comments on (\ref{Dhvgen}), (\ref{DT(2-z)/z}) 
towards gaining insight into ${\eta\over s}$~:

\noindent (1) As a consistency check, we see that for pure $AdS$ 
with $\theta=0$, $z=1$, we obtain $\mathcal{D}=\frac{1}{4\pi T}$~.
This corresponds to a relativistic CFT: the shear diffusion constant is
${\cal D} = {\eta\over \varepsilon + P}$ and thermodynamics gives 
$\varepsilon + P = Ts$, where $\varepsilon, P, s$ are energy, 
pressure and entropy densities. This gives the relation
$\frac{\eta}{s}=\mathcal{D}T$ and thereby\ 
$\frac{\eta}{s}=\frac{1}{4\pi}$~.

\noindent (2) Theories with metric (\ref{hvmetric}) and $\theta=0$ enjoy 
the Lifshitz scaling symmetry,\ $x_i\ra \lambda x_i,\ t\ra \lambda^z t$:\ 
Then the diffusion equation $\del_t j^t = D \del_i^2 j^t$ shows the 
diffusion constant to have scaling dimension $dim[{\cal D}]=z-2$,\ where 
momentum scaling is $[\del_i]=1$\ (or equivalently, $[x_i]=-1,\ [t]=-z$).
With temperature scaling as inverse time, we have $dim[T]=z$.
Thus on scaling grounds, the temperature scaling in (\ref{DT(2-z)/z}), 
which here is
\be\label{DT(2-z)/zLif}
{\cal D} = {1\over d-z+1} \Big({4\pi\over d+z-1}\Big)^{{z-2\over z}}\ 
T^{{z-2\over z}}\ ,
\ee
is expected, upto the $d,z$-dependent prefactors. For $z=2$, 
the diffusion equation (structurally like a Schrodinger equation) 
already saturates the scaling dimensions, and ${\cal D}$ has 
apparently no temperature dependence. As $T$ increases, ${\cal D}$ 
decreases for $z<2$: however ${\cal D}$ increases with $T$ for $z>2$.

Aspects of Lifshitz hydrodynamics have been studied in \eg\ 
\cite{Hoyos:2013qna}, \cite{Kiritsis:2015doa}. As discussed in 
\cite{Kiritsis:2015doa}, under the Lifshitz symmetry, we have the scalings\ 
$[T]=z ,\ [\varepsilon]=z+d-1 ,\ [P]=z+d-1 ,\ [s]=d-1 ,\ [\eta]=d-1$. 
Indeed for Lifshitz black branes with horizon $r_H$ and 
temperature (\ref{hvtemperature}), the entropy density is\ 
$s={r_H^{d-1}\over 4G_{d+1}} = 
{1\over 4G_{d+1}} ({4\pi\over d+z-1} T)^{{d-1\over z}}$. The 
thermodynamic relations give\ $\varepsilon+P=Ts$.\ The shear 
viscosity \cite{Kiritsis:2015doa} is\ 
$\eta={1\over 16\pi G_{d+1}} T^{{d-1\over z}}$\ satisfying the 
universal bound\
${\eta\over s} = {1\over 4\pi}$~.\ For this to arise from 
(\ref{DT(2-z)/zLif}), we guess that the relation between shear 
viscosity and the shear diffusion constant is
\be\label{etasDTLif}
{\eta\over s}\ =\ {(d-z+1)\over 4\pi} {\cal D} r_0^{2-z}\ 
=\ {(d-z+1)\over 4\pi} \Big({4\pi\over d+z-1}\Big)^{{2-z\over z}}\ 
{\cal D} T^{{2-z\over z}}\ .
\ee

\noindent (3) For $\theta\neq 0$, the scaling analysis of the Lifshitz
case is not applicable: however the temperature is
$\theta$-independent and the relation (\ref{DT(2-z)/z}) continues to
hold for generic $\theta$.\ Towards guessing the hydrodynamics from
the diffusion constant in this case, we first recall from
\cite{Kovtun:2003wp} that nonconformal branes give\ ${\cal D}={1\over
  4\pi T}$, and thereby ${\eta\over s}={1\over 4\pi}$ continues to
hold. On the other hand, \cite{Dong:2012se} observed that nonconformal
$Dp$-branes upon reducing on the sphere $S^{8-p}$ give rise to
hyperscaling violating theories with $z=1$ and $\theta\neq 0$. It
would therefore seem that the near-horizon diffusion analysis
continues to exhibit this universal behaviour since the sphere should
not affect these long-wavelength diffusive properties.\\
Happily, we see that (\ref{DT(2-z)/z}) for $z=1$ gives\
${\cal D} = {1\over 4\pi T}$, with the $\theta$-dependent prefactors
cancelling precisely. Thus all hyperscaling violating theories with
$z=1$ appear to satisfy the universal viscosity bound
\be
{\eta\over s} = {\cal D} T = {1\over 4\pi}\ .
\ee
Putting this alongwith the Lifshitz case motivates us to guess the 
universal relation
\be\label{eta/s,DT(2-z)/zHV}
{\eta\over s}\ =\ 
{(d-z-\theta+1)\over 4\pi} \mathcal{D} r_0^{2-z}\ =\ 
{(d-z-\theta+1)\over 4\pi} \Big({4\pi\over d+z-\theta-1}\Big)^{{2-z\over z}}\ 
\mathcal{D} T^{{2-z\over z}}\ =\ {1\over 4\pi}
\ee
between $\eta, s, {\cal D}, T$, for general exponents $z,\theta$.\ 
This reduces to (\ref{etasDTLif}) for the Lifshitz case $\theta=0$. 
One might wonder if the prefactors for $\theta\neq 0$ somehow conspire 
to violate the universal bound: in this regard, it is worth noting that 
$z,\theta$ appear in linear combinations in the prefactors. Alongwith 
the previous subcases, this suggests consistency of 
(\ref{eta/s,DT(2-z)/zHV}).\\
Finally we know that the entropy density is\ 
$s = {r_H^{d_{eff}}\over 4G_{d+1}} \sim {1\over 4G_{d+1}} T^{{d_i-\theta\over z}}$ 
in hyperscaling violating theories, with
$d_{eff}=d_i-\theta=d-1-\theta$ the effective spatial dimension. 
Then  (\ref{eta/s,DT(2-z)/zHV}) gives the shear viscosity as\ 
$\eta \sim {1\over 16\pi G_{d+1}} T^{{d_i-\theta\over z}}$.

It is fair to say that to study this conjecture in detail, it
is important to systematically understand the thermodynamic/hydrodynamic 
relations between the expansion of the energy-momentum tensor, the shear
viscosity $\eta$ and the diffusion constant ${\cal D}$. Towards this, 
it is worth putting the analysis here leading to (\ref{Dhvgen}), 
(\ref{DT(2-z)/z}), and the comments above in perspective with the 
calculation of viscosity via the Kubo formula 
$\eta = -\lim_{\omega\ra 0} {1\over\omega} {\rm Im} G^R_{xy,xy}(\omega)$, 
with $G^R$ the retarded Green's function \cite{Son:2002sd}\ (assuming 
$T_{ij}\sim \eta (\del_iv_j + \ldots)$ in the dual field theory). 
Holographically, this is obtained by modelling the $h^x_y$ perturbation 
as a massless scalar and thereby gleaning the 
$\langle T_{xy}T_{xy}\rangle$ correlation function: for various 
subfamilies in (\ref{hvmetric}), this has been carried out in \eg\ 
\cite{Pang:2009wa,Roychowdhury:2014lta,Kuang:2015mlf}. For instance 
in \cite{Kuang:2015mlf}, the appropriate solutions at zero momentum 
${\vec k}=0$ to the scalar wave equation in the near and far regions 
are matched to obtain 
$G^R = -i{\omega\over 16\pi G} {R^{d_i}\over r_{hv}^\theta} r_0^{d_i-\theta}$ 
and thereby $\eta$,\ where we have written the metric (\ref{hvmetric}) 
as\ $ds^2 = R^2 ({r\over r_{hv}})^{2\theta/d_i} (-f(r){dt^2\over r^{2z}} 
+ \ldots)$, explicitly retaining the dimensionful factors $R$ and 
the scale $r_{hv}$ inherent in these theories \cite{Dong:2012se}. 
Likewise the entropy density is\ 
$s={1\over 4G} {R^{d_i}\over r_{hv}^\theta} r_0^{d_i-\theta}$ 
from the area of the horizon, recovering ${\eta\over s}={1\over 4\pi}$ 
in agreement with our analysis 
(and $\theta$ cancels\footnote{We have seen that $\theta$ disappears 
from the temperature dependence of ${\cal D}$ in (\ref{DT(2-z)/z}). 
It would be inconsistent if $\theta$ remained, at least 
in cases where the hyperscaling violating phase arises from string 
constructions such as nonconformal branes which are known to have 
universal ${\eta\over s}$ behaviour, as discussed in comment (3) above.}).

We have seen the condition $z<2+d_i-\theta$ arising naturally from the
perturbations falling off asymptotically (\ref{At0soln}) in our
case. It is worth noting that although we implicitly regard
hyperscaling violating theories as infrared phases arising from
\eg\ string realizations in the ultraviolet, $z<2+d_i-\theta$ in the 
analysis here ensures that the ultraviolet structure of these
theories is essentially unimportant: the diffusion constant arises
solely from the near horizon long-wavelength modes. The theories
satisfying this condition are in some sense continuously connected to
$AdS$-like relativistic theories ($z=1, \theta=0$), as the analysis 
in Appendix \ref{appendix1} suggests. Identifying this condition 
from the point of view of holographic calculations appears more subtle.
While a detailed analysis is ongoing, we outline a few comments here, 
in part following discussions in \cite{Ross:2011gu} for Lifshitz 
theories. Bulk field modes have asymptotic fall-offs 
$\phi \sim r^{\Delta_-} (\phi_-+\ldots) + r^{\Delta_+} (\phi_++\ldots)$ 
near the boundary $r\ra r_c$, where $\Delta_-+\Delta_+= d_i-\theta+z$ 
\cite{Dong:2012se} (see also \cite{Chemissany:2014xsa,Taylor:2015glc}).
If $\Delta_-<\Delta_+$, the $\phi_-$ fall-off near the boundary 
$r\ra r_c\sim 0$ is slower (thus dominating), leading to fixed $\phi_-$ 
boundary conditions relevant for standard quantization ($\phi_-$ taken
as source). In particular, the momentum density operator ${\cal P}^i$
has dimension $d_i-\theta+1$ (while energy density has dimension
$d_i-\theta+z$): taking $\Delta_+=d_i-\theta+1$ gives $\Delta_-=z-1$,
so that $\Delta_-<\Delta_+$ implies $z<2+d_i-\theta$, which is
precisely our condition (\ref{genExp}). When this condition is
violated in a reasonable theory\footnote{unlike \eg\ $d_i=6, z=1,
\theta=9$, arising from the reduction of \eg\ D6-branes where the
asymptotics is ill-defined with gravity not decoupling.}, it would
seem that the analog of alternative quantization \cite{Klebanov:1999tb} 
will be applicable, with fixed $\phi_+$ boundary conditions. The case
$z=2+d_{eff}$ discussed in the remainder of the paper may be
interesting, with $\Delta_-=\Delta_+$. In these cases, $\theta$ may
not disappear. We hope to explore these issues further.

\subsection{Shear diffusion constant:\ \  $d-z-\theta=-1$\ \ (or\ 
$z=2+d_{eff}$)}

Now we consider the family of hyperscaling violating backgrounds 
(\ref{hvmetric}) with $d-z-\theta=-1$. 
In this case, the leading solution (\ref{At0soln}) for $A_t^{(0)}$ has 
logarithmic behaviour,
\be
A_t^{(0)} = C e^{-\Gamma t+iqx}\ \log\Big({r\over r_c}\Big)\ ,
\qquad\qquad d-z-\theta=-1\ .
\ee
Then working through, we have from \eqref{maindiffeqn}
\begin{equation}\label{violation}
\mathcal{D} = r_0^{d-\theta-1}\ \log\Big(\frac{1}{r_0r_c}\Big) = 
r_0^{z-2} \ \log\Big(\frac{1}{r_0r_c}\Big)\ .\ 
%\stackrel{r_0 \rightarrow 0}{\longrightarrow}\ 0
\end{equation}
This implies that in the low temperature limit $r_0\ra 0$, the diffusion 
constant becomes vanishingly small if $d_i-\theta>0$, or equivalently 
$z>2$.\
The energy conditions (\ref{energyconds}) eliminating $\theta=d-z+1$ give\
$(z-2) \big((d-1)z - 2d + z\big) \geq 0 ,\ \ (z-1)(z-2) \geq 0$.\
When $\theta=0$, we obtain Lifshitz theories: the energy
conditions become\ $(z-1)(d-1+z) \geq 0$. Then the condition here
is\ $z=d+1=2+d_i$. Since we are considering theories in $d+1\geq 3$ 
bulk dimensions, $z\geq 3$ consistent with the energy conditions above.\
With $\theta\neq 0$, we have\ $z = 2 + d_{eff}$. In \cite{Dong:2012se}, 
it was noted that the entropy scaling\ $S\sim T^{(d_i-\theta)/z}$ implies
that the specific heat is positive if ${d_i-\theta\over z}\geq 0$. 
Here we have $S\sim T^{d_{eff}\over 2+d_{eff}}$ so that positivity of the 
specific heat gives ${d_{eff}\over 2+d_{eff}}>0$ if $d_{eff}>0$. 
Relatedly, we recall that entanglement entropy has novel scaling 
behaviour in the window $d_i-1\leq \theta \leq d_i$, which does not 
involve $z$. The entangling surface has been observed to have some 
instabilities for $\theta>d_i$. The present condition  $z=2+d_{eff}$ 
is thus distinct from and compatible with them.

We show in Appendix \ref{analysisspecial} that the conditions 
(\ref{combinedcriteria}) on the stretched horizon now become
\be\label{specialbound}
\exp \Big( -\frac{T^{2/z}}{q^2}\frac{1}{\log \frac{1}{r_0r_c}}\Big)\ \ll\ 
\frac{\frac{1}{r_0}-r_{h}}{\frac{1}{r_0}}\ \ll\ 
\frac{q^2}{T^{2/z}}\log^2 \frac{1}{r_0r_c}\ .
\ee
In the generic cases (\ref{combinedcriteria}), the power law behaviour
ensured that the short distance cutoff decoupled from the near-horizon
behaviour (\eg\ since ${1\over r_0^\#} \gg r_c^\#$). Here the solution
$A_t^{(0)}$ contains a logarithm which requires a scale, which filters
through to (\ref{specialbound}).  While it is unusual for the UV
cutoff $r_c$ to appear in what is manifestly a hydrodynamic or
long-wavelength regime that we have restricted our analysis to,
(\ref{combinedcriteria}) implies (\ref{specialbound}) since $r_c\ll
{1\over r_0}$ implies $\log {1\over r_0r_c}\gg 1$ so that the window
for the stretched horizon is not over-constrained. However, the
subleading corrections (\ref{delrAt1}) to the gauge field
perturbations again contain terms involving $\log {1\over r_0r_c}$
factors affecting the validity of the series expansion in ${q^2\over
T^{2/z}}$.

It appears reasonable to conclude that the series expansion is perhaps
breaking down in this case. Towards gaining some insight into this, it
is useful to look for gauge/string embeddings of these effective
gravity theories. In this regard, we recall that $AdS$ plane waves
(equivalently, highly boosted black branes) upon $x^+$-reduction give
rise to hyperscaling violating spacetimes with certain values for the
$z,\theta$-exponents \cite{Narayan:2012hk,Singh:2012un,Narayan:2012wn}. 
It turns out that the exponents satisfy $d-z-\theta=-1$. We will
discuss this in what follows.

\section{Diffusion constant: highly boosted black branes}

A simple subclass of (zero temperature) hyperscaling violating theories 
can be constructed from the dimensional reduction of $AdS_{d+2}$ plane 
wave spacetimes \cite{Narayan:2012hk,Narayan:2012wn}
\bea\label{adspw}
ds^2 &=& {R^2\over r^2} [-2dx^+dx^- + dx_i^2 + dr^2] + R^2Qr^{d-1} (dx^+)^2 
+ R^2 d\Omega_S^2
\quad \longrightarrow\\
ds^2 &=& r^{2 \theta \over d_i} \Big(-{dt^2 \over r^{2z}} + 
{\sum_{i=1}^{d_i}  dx_i^2 + dr^2 \over r^2 }\Big), \quad\ \ 
z={d+3\over 2} ,\quad \theta={d-1\over 2}\ ,\quad d_i=d-1 .\ \  
\label{hypviol}
\eea
%$ds^2 =\frac{R^2}{r^2}(-2dx^{+}dx^{-}+dx_i^2+dr^2)+g_{++}(r)(dx^{+})^2$. 
These can be obtained from a low-temperature, large boost limit
 \cite{Singh:2012un} of boosted black branes \cite{Maldacena:2008wh} 
arising from the near horizon limits of the conformal $D3$-, $M2$- 
and $M5$-branes. Similar features arise from reductions of 
nonconformal $Dp$-brane plane waves \cite{Narayan:2013qga} 
\cite{Singh:2012un}, with exponents
\be\label{hvNonconf}
z = {2(p-6)\over p-5}\ ,\qquad \theta = {p^2-6p+7\over p-5}\ ,\qquad
d_i=p-1\ ,
\ee
where the $Dp$-brane theory after dimensional reduction on the 
sphere $S^{8-p}$ and the $x^+$-direction has bulk spacetime dimension 
$d+1\equiv p+1$. The holographic entanglement 
entropy in these theories exhibits interesting scaling behaviour 
\cite{Narayan:2012ks,Narayan:2013qga,Mukherjee:2014gia,Narayan:2015lka}.

To obtain the finite temperature theory, let us for simplicity consider 
the $AdS_5$ black brane 
\begin{equation}\label{blackbrane}
ds^2 = \frac{R^2}{r^2}\Big(-(1-r_{0}^4r^4)dt^2+dx_3^2 + 
\sum_{i=1}^{2}dx_i^2\Big)+R^2 \frac{dr^2}{r^2(1-r_0^4r^4)}\ ,
\end{equation}
which is a solution to the action $S = \frac{1}{16 \pi G_5}\int d^5x
\sqrt{-g^{(5)}}(R^{(5)}-2\Lambda)$.  Rewriting \eqref{blackbrane} in
lightcone coordinates and boosting as $x^{\pm}\rightarrow
\lambda^{\pm}x^{\pm}$, we obtain
\be
ds^2 = {R^2\over r^2} \Big( -2dx^+dx^-+ {r_0^4 r^4\over 2}
(\lambda dx^++\lambda^{-1} dx^-)^2 +\sum_{i=1}^2 dx_i^2\Big)
+ {R^2 dr^2\over r^2 (1-r_0^4r^4)}\ .
\ee
Writing in Kaluza-Klein form
\begin{equation}\label{upstairsKK}
ds^{2}=\frac{R^{2}}{r^{2}}\left[-\frac{(1-r_{\tinynot}^{4}r^{4})}{Qr^{4}}(dx^{-})^{2}+dx^{2}+dy^{2}+\frac{dr^{2}}{(1-r_{\tinynot}^{4}r^{4})} \right] + QR^{2}r^{2}
\left(dx^{+}-\frac{(1-\frac{r_{\tinynot}^{4}r^{4}}{2})}{Qr^{4}}dx^{-}\right)^2\ ,
\end{equation}
where $Q=\dfrac{\lambda^{2}r_{\tinynot}^{4}}{2}$ and compactifying along 
the $x^{+}$ direction gives
\begin{equation}\label{thermalhv}
ds^{2}=(Q^{1/2}R^3) r \Big[-\frac{(1-r_{\tinynot}^{4}r^{4})}{Qr^{6}}(dx^{-})^{2}
+ {dx^{2}+dy^{2}\over r^2} + \frac{dr^{2}}{r^2 (1-r_{\tinynot}^{4}r^{4})}\Big] .
\end{equation}
This is simply the hyperscaling violating metric (\ref{hvmetric}) 
with $z=3,\ \theta=1,\ d_i=2$, in \cite{Narayan:2012hk}, but now at 
finite temperature. It is a solution to the equations stemming from the 
4-dim Einstein-Maxwell-Dilaton action\ 
$S = \frac{1}{16 \pi G_4} \int d^4 x \sqrt{-g^{(4)}}\ 
(R^{(4)} -2\Lambda e^{-\phi}-\frac{3}{2}g^{\mu \nu}\partial_{\mu}\phi
\partial_{\nu}\phi -\frac{1}{4}e^{3\phi}F^{\mu \nu}F_{\mu \nu} )$\ 
which arises upon dimensional reduction along the $x^+$-direction of 
the 5-dim Einstein action\ 
$S = \frac{1}{16 \pi G_5}\int d^5x \sqrt{-g^{(5)}}(R^{(5)}-2\Lambda)$. The
scalar field has the profile $e^{2\phi} = R^2Qr^2$ while the gauge field
is $\mathcal{A}_t=-\frac{1+f}{2Qr^4} ,\ \mathcal{A}_i =0$ with
$f=1-r_0^4r^4$. The finite temperature theory is of course obtained 
by taking the boost $\lambda$ to be large but finite, and the temperature 
$r_0$ to be small but nonzero, while holding $Q={\lambda^2 r_0^4\over 2}$ 
fixed. The boost simply serves to create a hierarchy of scales in 
the energy momentum components\ $T_{++}\sim \lambda^2 r_0^4\sim Q ,\ 
T_{--}\sim {r_0^4\over\lambda^2}\sim {r_0^8\over Q}~,
\ T_{+-}\sim r_0^4 ,\ T_{ij}\sim r_0^4\delta_{ij}$,\ 
while keeping them nonzero.

The $z,\theta$-exponents (\ref{hypviol}) arising in these reductions
satisfy $d-z-\theta=-1$, coinciding with the special case discussed
earlier. This is also true for nonconformal $Dp$-brane plane waves
(\ref{hvNonconf}). It is worth noting that this relation between the
exponents is distinct from the window $d_i-1\leq \theta\leq d_i$ where
the holographic entanglement entropy exhibits novel scaling behaviour:
in particular the present relation involves the Lifshitz exponent. The
diffusion constant for this class of hyperscaling violating theories
then has the logarithmic behaviour \eqref{violation} described
earlier, provided we restrict to modes that describe the lower
dimensional theory.

%\subsection{Perturbations in boosted black branes}

In the above $x^+$-compactification, we see that $x^-$ above maps to
the time coordinate $t$ below. Thus mapping the
perturbations between the higher dimensional description and the
hyperscaling violating one, we see that the metric in KK-form
(\ref{upstairsKK}) including the shear gravitational perturbations is
of the form (with gauge condition ${\tilde h}_{\mu r}, {\tilde h}_{rr}=0$)
\be\label{KKformPert}
ds^2 = {\tilde g}_{--} (dx^-)^2 + {\tilde g}_{ii} dx_i^2 
+ {\tilde g}_{rr} dr^2 + 2{\tilde h}_{-y} dx^- dy + 2{\tilde h}_{xy} dxdy 
+ {\tilde g}_{++} (dx^+ + A_-dx^-)^2\ ,
\ee
where $A_-$ is the background gauge field in the lower dimensional 
description. In other words, the perturbations map as\ 
${\tilde h}_{-y} \ra h_{ty} ,\ {\tilde h}_{xy} \ra h_{xy}$,\ upto the 
conformal factor arising from the $x^+$-reduction. In addition, the 
$x^+$-reduction requires that the perturbations ${\tilde h}_{-y}, 
{\tilde h}_{xy}$ are $x^+$-independent. This in turn translates to the 
statement that the near horizon diffusive modes are of the form
\be
h_{\mu y}(r) e^{-k_-x^-+ik_xx}\ , \qquad k_+=0\ , 
\qquad\qquad[\mu=x^-, x]\ ,
%h_{\mu y}(r) e^{ik_+x^++ik_-x^-+ik_xx} \ra 
\ee
\ie\ the nontrivial dynamics in the lower dimensional description 
arises entirely from the zero mode sector $k_+=0$ of the full theory.

Likewise, vector perturbations $\delta A_t, \delta A_y$ in the lower 
dimensional theory arise in (\ref{KKformPert}) as
\bc
$\ldots + g_{++} (dx^+ + A_-dx^- + {\tilde h}_{+-} dx^- + {\tilde h}_{+y} dy)^2$.
\ec
We see that these arise from gravitational perturbations 
${\tilde h}_{+-}, {\tilde h}_{+y}$.

To ensure that the massive KK-modes from the $x^+$-reduction decouple
from these perturbations, it suffices to take the $x^+$-circle size
$L_+$ to be small relative to the scale set by the horizon, \ie\
$L_+\ll {1\over r_0}$~: equivalently, the temperature is small
compared to the KK-scale ${1\over L_+}$~. The ultraviolet cutoff near
the boundary is $r_c\sim Q^{-1/4}\ll {1\over r_0} $: the hyperscaling
violating phase is valid for $r\gtrsim Q^{-1/4}$.

Finally to map \eqref{thermalhv} to \eqref{hvmetric} precisely, we 
absorb the factors of the energy scale $Q$ by redefining\
$\tilde{x}^{-} = \frac{x^{-}}{\sqrt{Q}}$.\ Now the shear diffusion 
constant can be studied as in the hyperscaling violating
theory previously discussed, by mapping it to charge diffusion in an
auxiliary theory obtained from the finite temperature
$x^+$-compactified theory by compactifying along say the
$y$-direction.  This requires mapping the shear gravitational
perturbations to the lower dimensional auxiliary gauge fields as\
$A_t\propto {\tilde h}_{-y} ,\ A_y \propto {\tilde h}_{xy}$, which 
can then be set up in a series expansion in the near horizon region. 
Thus finally the shear diffusion constant follows from 
(\ref{violation}) giving\ ${\cal D} = r_0 \log ({1\over r_0 Q^{-1/4}})$.\ \
For $Q$ fixed, as appropriate for the lower dimensional theory, we see 
that the low temperature limit $r_0\ra 0$ gives a vanishing shear 
diffusion constant suggesting a violation of the viscosity bound.
It is worth noting that the diffusion equation here is\ 
$\del_{{\tilde x}^-} j^- = {\tilde {\cal D}} \del_i^2 j^-$ where 
${\tilde x}^-={x^-\over\sqrt{Q}}$ reflecting the Lifshitz 
exponent $z=3$.

Noting that $Q\sim \lambda^2 r_0^4$, the diffusion equation and 
constant in the upstairs theory are\ 
\be\label{DlogL/L}
\del_{x^-} j^- \sim {{\tilde {\cal D}}\over \sqrt{Q}} \del_i^2 j^-\ ,
\qquad\qquad 
{\cal D} r_0 \sim {{\tilde {\cal D}}\over\sqrt{Q}} r_0 
\sim {1\over \lambda} \log\lambda\ .
\ee
The $r_0\ra 0$ limit of the lower dimensional theory (where $T\sim r_0^3$)
implies a highly boosted limit $\lambda\ra\infty$ of the black brane 
for fixed $Q$: here ${\cal D} r_0$ vanishes. However this appears to 
be a subtle limit of hydrodynamics. 
From the point of view of the upstairs theory of the unboosted black
brane, shear gravitational modes are $h_{ty}, h_{xy}$. Upon boosting,
it would appear that these mix with other perturbation modes as well,
suggesting some mixing between shear and bulk viscosity. From the
point of view of the boosted frame, this system has anisotropy
generated by the boost direction. Previous studies of anisotropic 
systems and shear viscosity include \eg\ \cite{Mateos:2011tv,
Erdmenger:2011tj,Rebhan:2011vd,Polchinski:2012nh,Mamo:2012sy,
Jain:2014vka,Jain:2015txa}. (See also \eg\ \cite{Cremonini:2011iq} 
for a review of the viscosity bound.)\
In the present case, the shear viscosity tensor can be analysed from a
systematic study of the expansion of the energy-momentum tensor of the
finite temperature Yang-Mills fluid in the highly boosted
regime. However the scaling (\ref{DlogL/L}) is likely to be realized
only after phrasing the boosted black brane theory in terms of the
variables appropriate for the lower dimensional hyperscaling violating
theory (which arises in the $k_+=0$ subsector as discussed above). 
It would be interesting to understand the hydrodynamics in
the lower dimensional theory better, as a null reduction of the boosted
black brane theory, perhaps similar in spirit to nonconformal brane 
hydrodynamics \cite{Kanitscheider:2008kd,Kanitscheider:2009as} 
as a reduction of nonlinear hydrodynamics \cite{Bhattacharyya:2008jc} 
of black branes in M-theory. We hope to explore this further.

\vspace{7mm}
%\newpage

{\footnotesize \noindent {\bf Acknowledgements:}\ \
It is a pleasure to thank S. Govindarajan, D. Jatkar, A. Laddha, 
R. Loganayagam, S. Minwalla, B. Sathiapalan and A. Sen for discussions 
on this work. We thank the string theory group, HRI, Allahabad, for 
hospitality while this work was in progress. This work is partially 
supported by a grant to CMI from the Infosys Foundation.
}

%\vspace{-3mm}

\appendix

\section{Diffusion analysis details:\ generic case}\label{appendix1}

%\subsection*{Field Equations}

In the near-horizon region, the metric (\ref{hvmetric}) is approximated 
as\ 
$g_{tt}(r) \approx -\gamma_{0}( \frac{1}{r_0}-r ) , \ 
g_{rr}(r) \approx \frac{\gamma_{r}}{( \frac{1}{r_0}-r )} ,\ 
g_{xx}(r) \approx const$, for constants $\gamma_0, \gamma_r$ :\ the 
Maxwell equations (\ref{gaugefieldaction}) simplify substantially here, 
as a series expansion in ${q^2\over T^{2/z}}$ for the gauge fields. 
Here we only mention the modifications in the analysis of 
\cite{Kovtun:2003wp} arising in the present context. 
An intermediate step in the self-consistent analysis gives
\begin{equation}\label{estimatea}
F_{tr} \sim r_0^{2(z-1)}\cdot\frac{q}{\Gamma^2}\cdot 
\frac{(1/r_0)-r_h}{1/r_0} \partial_r F_{tx}\ ,
\end{equation}
which is then used to obtain a wave equation for $F_{tx}$: we choose 
the solution that is ingoing at the horizon, and then solve for the 
various gauge field components. Self-consistency constrains the 
location of the stretched horizon
\begin{equation}
\label{criteriaa}
\frac{(1/r_0)-r_h}{1/r_0} \ll \frac{1}{r_0^{2(z-1)}}\cdot
\frac{\Gamma^2}{q^2}\ .
\end{equation}
For thermal $AdS$ ($z=1$), this becomes
$\frac{(1/r_0)-r_h}{1/r_0} \ll \frac{\Gamma^2}{q^2}$ 
as in \cite{Kovtun:2003wp}.

With small but nonzero $q$, we write $A_t, A_x$ as series expansions 
\eqref{ansatz1}, (\ref{axsoln}), in $\frac{q^2}{T^{2/z}}$ and impose
(\ref{ansatz2}) and the boundary conditions that $A_t, A_x\ra 0$ at 
$r=r_c\sim 0$. 
Using the $A_t^{(0)}, A_x^{(0)}$ solutions (\ref{ansatz1}), (\ref{axsoln}), 
(\ref{At0soln2}), (\ref{Ax0soln}), in the near horizon region 
$\frac{1}{r_0}-r \ll \frac{1}{r_0}$ shows\ $\frac{A_x^{(0)}}{A_t^{(0)}}
\sim \frac{1}{r_0^{2(z-1)}}\frac{\Gamma}{q}\log (\frac{1/r_0}{1/r_0-r})$. 
Thus (\ref{ansatz2}) is valid only if
\begin{equation}\label{criteriab}
\frac{1/r_0}{(1/r_0)-r_h} \ll e^{\frac{q^2r_0^{2(z-1)}}{\Gamma^2}}\ .
\end{equation}
Combining \eqref{criteriaa}, \eqref{criteriab}, gives the window 
\begin{equation}
\label{combinedcriteriaA}
e^{-\frac{q^2r_0^{2(z-1)}}{\Gamma^2}} \ll \frac{(1/r_0)-r_h}{1/r_0}
\ll \frac{1}{r_0^{2(z-1)}} \frac{\Gamma^2}{q^2}\ ,
\end{equation}
for the stretched horizon $r_h$ in terms of the perturbation 
parameters ${\Gamma^2\over q^2}\ll 1$.

%\subsection*{Fick's Law}\label{appendix3}

We define the following gauge field currents on the stretched horizon
\begin{equation}
j^{x}=n_rF^{xr}=-\frac{F_{tx}}{g_{xx}\sqrt{-g_{tt}}}\ , \quad \qquad 
j^{t}=n_rF^{tr}=-\frac{1}{g_{tt}\sqrt{g_{rr}}}F_{tr}\ .
\end{equation}
The assumption (\ref{ansatz2}) implies
we can approximate the field strength $F_{tx} \approx -\partial_{x}A_{t}$,\
and $\partial_{r}F_{tx}\approx \partial_{x}F_{tr}$\ in radial gauge. 
This gives\
$\partial_x j^{t} = -\frac{1}{g_{tt}\sqrt{g_{rr}}}\partial_{x} F_{tr} =
-\frac{1}{g_{tt}\sqrt{g_{rr}}}\partial_r F_{tx}$. Further, multiplying
and dividing by $F_{tx}$ and using the definition of $j^{x}$, we get
$\partial_x j^{t} =
-\frac{g_{xx}}{\sqrt{-g_{tt}g_{rr}}}\frac{\partial_rF_{tx}}{F_{tx}}j^{x}$,\ 
\ie\
%This approximation simplifies the form of $j^{x}$ and can be written as 
%a time derivative of the charge density $j^{t}$, \ie\ 
\be\label{FickA}
j^{x} =-
\frac{\sqrt{-g_{tt}g_{rr}}}{g_{xx}}\frac{F_{tx}}{\partial_r F_{tx}}
\partial_x j^{t}\ \equiv\ -D \partial_x j^t\ ,
\ee
which is Fick's Law\footnote{The 
antisymmetry of $F_{\mu \nu}$ implies $n_{\mu}j^{\mu}=0$, \ie\ 
the current is parallel to the stretched horizon (only $n_r$ is 
nonzero with $g^{rr}n_r^2 = 1$). Contracting gives\ 
$n_{\nu}\partial_{\mu}(\frac{\sqrt{-g}}{\geff^2}F^{\mu \nu})=0
\Rightarrow n_r(\partial_{M}F^{M r})=0$, with $M=t,x_i$. This gives
current conservation\ $\partial_{\mu}j^{\mu} 
= \partial_{M}(n_{\nu}F^{M\nu})=n_r(\partial_MF^{Mr})=0$.}.\ 
From the solutions to $A_{t}$ and $A_{x}$, we have
\begin{equation}\label{At/Ftr}
\left. -\frac{F_{tx}}{\partial_r F_{tx}}\right|_{r\approx r_h} 
\approx \left. -\frac{A_t}{F_{tr}} \right|_{r \approx r_h}
= \frac{\sqrt{-g(r_h)}}{\geff^2(r_h) g_{tt}(r_h)g_{rr}(r_h)}
\int_{r_c}^{r_h}\frac{g_{tt}(r')g_{rr}(r')\geff^2(r')}{\sqrt{-g(r')}}dr'\ .
\end{equation}
Thus from Fick's Law (\ref{FickA}), we read off the shear diffusion 
constant 
\begin{equation}
\begin{aligned}
\label{chargediff2}
D &= \frac{\sqrt{-g_{tt}(r_h)g_{rr}(r_h)}}{g_{xx}(r_h)}\cdot \left.
\frac{A_t}{F_{tr}}\right|_{r=r_h}\ ,
\end{aligned}
\end{equation}
evaluated at the stretched horizon $r_h$, where the boundary is 
$r_c\ll r_h$~. Using (\ref{At/Ftr}) gives (\ref{chargediff}).

For generic exponents, the diffusion constant (\ref{Dhvgen}) becomes 
$D \sim r_0^{z-2} \sim T^{(z-2)/z}$. On the other hand, the diffusion 
equation gives\ $\Gamma = Dq^2$. These give the condition
\begin{equation}\label{fickcriteria}
\frac{\Gamma}{q} \sim \frac{q}{T^{\frac{2}{z}-1}}\ .
\end{equation}
Using this estimate, \eqref{combinedcriteriaA} becomes 
(\ref{combinedcriteria}), which is always satisfied for sufficiently 
small $\frac{q^2}{T^{2/z}}$. In particular for thermal $AdS$ we have 
$z=1,\ D \sim T^{-1}$ so (\ref{combinedcriteria}) becomes\ 
$e^{-\frac{T^2}{q^2}} \ll \frac{(1/r_0)-r_h}{1/r_0} \ll 
\frac{q^2}{T^2}$~, the condition obtained in \cite{Kovtun:2003wp}.

Using the expansion over $q^2$ for $A_t$ and $A_x$ in the gauge field 
equations, it can be checked that the leading $O(q^0)$ terms are 
consistent with the ansatz for $A_t^{(0)}, A_x^{(0)}$ in the regime 
\eqref{combinedcriteria}. Likewise the subleading terms can be 
evaluated: collecting terms of $O(q^2)$ consistently, using 
\eqref{fickcriteria} and simplifying gives
\begin{equation}
\partial_r A_t^{(1)} \sim r_0 \Big[\frac{q^2}{T^{2/z}}\log
\Big(\frac{1/r_0}{(1/r_0)-r}\Big) + \frac{q^4}{T^{4/z}}
\log^2 \Big(\frac{1/r_0}{(1/r_0)-r}\Big) \Big] A_t^{(0)}\ .
\end{equation}
Using \eqref{combinedcriteria}, we see that 
$\partial_r A_t^{(1)} \ll A_t^{(0)}$, and after integrating, that 
$A_t^{(1)} \ll A_t^{(0)}$, verifying that these are indeed subleading.\
Likewise, we find
\begin{equation}
A_x^{(1)}  \sim \Big[\frac{q^2}{T^{2/z}}\log\Big( \frac{1/r_0}{1/r_0 -r}\Big) 
+ \frac{q^4}{T^{4/z}}\log ^2\Big( \frac{1/r_0}{1/r_0 -r}\Big) \Big] A_x^{(0)}
\ll A_x^{(0)}\ .
\end{equation}

\section{Diffusion analysis details:\ special case}\label{analysisspecial}

For the case $d-z-\theta=-1$,  we obtain\
$\frac{A_x^{(0)}}{A_t^{(0)}} \sim \frac{1}{r_0^{2(z-1)}}\frac{\Gamma}{q}
\frac{\log (\frac{1/r_0}{1/r_0-r})}{\log (\frac{1}{r_0r_c})}$ so that 
imposing \eqref{ansatz2} gives 
\begin{equation}
\frac{1}{r_0^{2(z-1)}}\cdot \frac{\Gamma^2}{q^2}\cdot 
\frac{\log (\frac{1/r_0}{1/r_0 -r})}{\log (\frac{1}{r_0r_c})} \ll 1\ .
\end{equation}
From the estimates obtained for ${\cal D}$ from the diffusion equation 
and the diffusion integral, we obtain\ $\frac{\Gamma}{q}
\sim \frac{q}{T^{\frac{2}{z}-1}}\log (\frac{1}{r_0r_c})$ instead of 
\eqref{fickcriteria}. Using this in \eqref{criteriaa}, we obtain the 
modified bound\
$\frac{(1/r_0)-r_h}{1/r_0}\ll 
\frac{q^2}{T^{2/z}}\log^2 (\frac{1}{r_0r_c})$.\ Likewise, 
\eqref{criteriab} also changes to\ 
$\frac{q^2}{T^{2/z}} \log (\frac{1/r_0}{(1/r_0)-r_h})\log
\frac{1}{r_0r_c} \ll 1$.\ 
The subleading terms now give
\begin{equation}\label{delrAt1}
\partial_r A_t^{(1)} \sim r_0 \Big[\frac{q^2}{T^{2/z}}\log
\Big(\frac{1/r_0}{(1/r_0)-r}\Big) + \frac{q^4}{T^{4/z}}\log^2
\Big(\frac{1/r_0}{(1/r_0)-r}\Big) \log \Big({1\over r_0r_c}
\Big)\Big]A_t^{(0)}\ .
\end{equation}
Within the regime \eqref{specialbound}, it would appear that 
$\partial_r A_t^{(1)} \ll A_t^{(0)}$: however $r_0r_c\ll 1$ implies that 
$\log ({1\over r_0r_c})$ is large so that the $O(q^4)$ term need not be 
small even if ${q^2\over T^{2/z}}\ll 1$, suggesting a breakdown of the 
series expansion.

\end{document}